\documentclass[aps,prb,twocolumn,superscriptaddress]{revtex4}
\usepackage[icelandic]{babel}
\usepackage{times,float}
\usepackage{t1enc}
\usepackage[latin1]{inputenc}
\usepackage{amsmath}
\usepackage{amssymb}
\usepackage{graphicx}

\usepackage{color}
\usepackage{bm}
\usepackage{epstopdf}

\newcommand{\LDQ}{,,}          
\newcommand{\RDQ}{``{ }}               

\begin{document}
\selectlanguage{icelandic}

\author{E.B. Magnusson}
\affiliation{Raunvísindastofnun Háskólans, Háskóli Íslands, Dunhagi 3,
IS-107, Reykjavik, Iceland}

\author{I. A. Shelykh}
\affiliation{Raunvísindastofnun Háskólans, Háskóli Íslands, Dunhagi 3,
IS-107, Reykjavik, Iceland}
\affiliation{International Institute of
Physics, Av. Odilon Gomes de Lima, 1772, Capim Macio, 59078-400,
Natal, Brazil}

\date{\today}
\title{Brotaskömmtun skotleiðninnar í einvíðum rafeinda og holukerfum\\
Fractional quantization of the ballistic conductance in electron and
hole systems}

\begin{abstract}
Í grein þessari verður skömmtun skotleiðninnar í einvíðum kerfum
útskýrð og farið yfir nýjustu og helstu fræðilíkön á sviði \LDQ 0,7
fráviksins \RDQ og tengdra fyrirbrigða í einvíðum rafeinda og
holukerfum. Við kynnum hugmynd okkar um að brotaskömmtun
skotleiðninnar leiði af Heisenberg skiptaverkun milli staðbundins
rafbera í skammtapunktsnertunni og frjálsra leiðnirafbera og sýnum
fram á að hegðun leiðninnar er eigindlega ólík milli rafeinda og
holukerfa.

In the present article we give a brief overview of the theories of
the 0.7 anomaly of the ballistic conductance. The special attantion
is given to the concept of a fractional quantization of the
ballistic conductance arising from exchange interaction of the
Heisenberg type between the carrier localized in the region of the
quantum point contact and freely propagating carriers in electron
and hole systems.
\end{abstract}

\maketitle

\section{Inngangur}
Miðsæ eðlisfræði er með þeim sviðum þétteðlisfræðinnar sem eru í hvað skjótastri þróun. Þökk sé framförum í örtækni varð fyrir um tveimur áratugum hægt að búa til í rannsóknastofum nokkuð sem nálgast að vera einvíðar rásir. Í slíkum rásum er hreyfing rafberans heft í tveimur víddum en óheft í þeirri þriðju. Rafflutningurinn getur verið ýmist sveimkenndur eða skotkenndur, eftir því hvert hlutfallið er milli lengd rásar og meðalferðavegalengd milli ófjaðrandi árekstra. Í skotkenndum flutningi verða allir ófjaðrandi árekstrar og varmatapið sem þeim fylgja, í tvívíðu stórsæju tengjum rásarinnar en ekki rásinni sjálfri. Í þessari nálgun er leiðnin í hlutfalli við leiðniskammtinn $G_0=2e^2/h$ \cite{Landauer, Buttiker}.

\begin{equation}
G=2\frac{e^{2}}{h}N,  \label{LBformula}
\end{equation}
þar sem $N$ er fjöldi opinna leiðniástanda. Honum er hægt að breyta með því að stilla hliðspennuna $V_g$ og skoða þannig þrepahegðun skotleiðninnar \cite{Wees,Wharam}.

\subsection{Skotleiðniskömmtunin}
\begin{figure}[tbp]
\includegraphics[width=1.0\columnwidth,keepaspectratio]{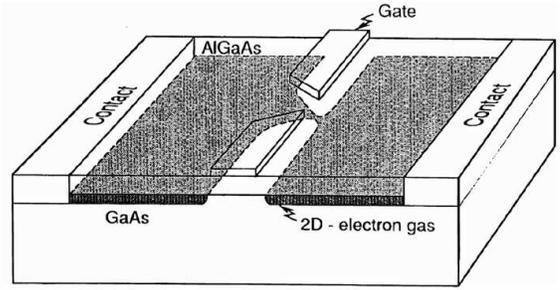}
\caption{Skammtapunktsnerta með tvívíðum stórsæjum tengjum. Hliðið býr til lítið svæði sem verður nánast einvítt eða nógu mjótt til að aðgreina orkuborðana.}
\label{FigQPC}
\end{figure}

\begin{figure}[tbp]
\includegraphics[width=1.0\columnwidth,keepaspectratio]{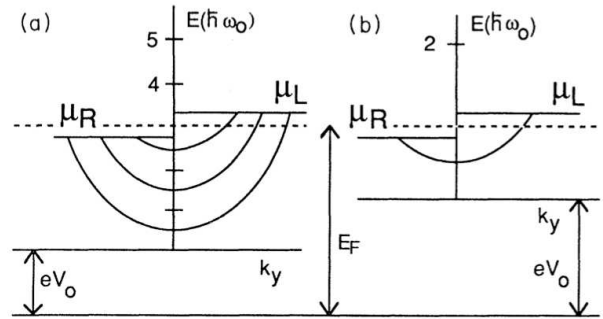}
\caption{Útskýringamynd fyrir orkuborða í nærri einvíðu kerfi. Hægri: Einungis einn borði til staðar. Vinstri: Margir borðar tilkomnir vegna endanlegrar breiddar vírs.}
\label{Figsubbands}
\end{figure}

Til að útskýra nánar skömmtun skotleiðninnar skulum við reikna lauslega leiðnina fyrir kerfið á mynd \ref{FigQPC}.
Straumur um skammtapunktsnertu eða vír af lengd $L$ og með einn opinn orkuborða $\nu$ (sjá mynd \ref{Figsubbands}) nálægt Fermiorkunni má skrifa sem \cite{DattaAtT}

\begin{align}
I&=\frac{-e}{L} \sum_{v_x(k_x)>0} v_x(k_x) = \frac{-e}{L} \sum_{v_x(k_x)>0} \frac{1}{\hbar}\frac{\partial E_\nu (k_x)}{\partial k_x}\\ \nonumber
&= -e \int \frac{\text d k_x}{2\pi} \frac{1}{\hbar}\frac{\partial E_\nu (k_x)}{\partial k_x}= \frac{-e}{h} \int \text d E_\nu
\end{align}
þar sem summað er yfir setin ástönd með jákvæðan grúppuhraða sem ekki styttast út af setnu ástandi með jafnstóran neikvæðan hraða.
Ef engin spennumunur er milli enda sýnisins flæðir enginn straumur, því ástönd með neikvæðan hraða eru jafnt setin og ástöndin með jákvæðan hraða. Ef hins vegar spennumunurinn $V$ er til staðar, þá verða ástöndin ójafnt setin á orkubilinu $E_F\pm eV/2=\mu_1,\mu_2$. Ef við hugsum okkur þá að á þessu bili séu einungis ástöndin með jákvæðan hraða setin, þá fæst að straumurinn er

\begin{equation}
I=\frac{-e}{h}(\mu_1-\mu_2)=\frac{e^2}{h}V
\end{equation}
þ.a. leiðnin er

\begin{equation}
G=I/V=\frac{e^2}{h}
\end{equation}
Ef fleiri en einn orkuborði er til umráða eiga sömu útreikningar við þá alla, og við leggjum saman niðurstöðurnar frá hverjum. Sér í lagi er í mörgum efnum án utanaðkomandi segulsviðs spunamargfeldni, þ.e. tveir eins borðar eru til staðar, einn fyrir hvorn spunann. Þar af leiðir skilgreiningin á leiðniskammtinum, en hann á við leiðni gegnum eitt opið ástand með spunamargfeldni:

\begin{equation}
G_0=\frac{2e^2}{h}
\end{equation}

\begin{figure}[tbp]
\includegraphics[width=1.0\columnwidth,keepaspectratio]{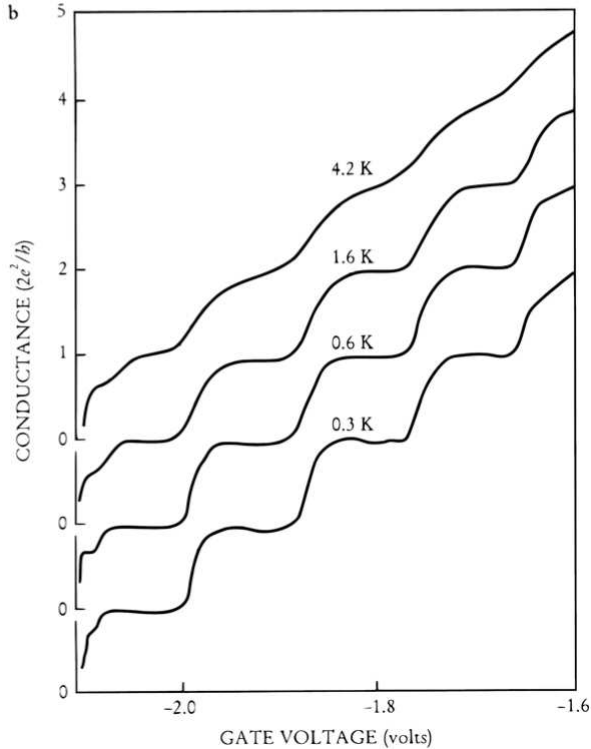}
\caption{Leiðniþrepamunstur sem fæst með því að auka smám saman hliðspennuna þannig að fleiri og fleiri orkuborðar taki þátt í leiðninni.}
\label{Figstairs}
\end{figure}

Ef hliðspennan er smám saman aukin þannig að farið er frá því að enginn orkuborði taki þátt í leiðninni, yfir í einn, þá tvo og koll af kolli, fæst þrepamynd eins og sést á mynd \ref{Figstairs}.

\subsection{\LDQ 0,7 frávikið\RDQ}
\begin{figure}[tbp]
\includegraphics[width=1.0\columnwidth,keepaspectratio]{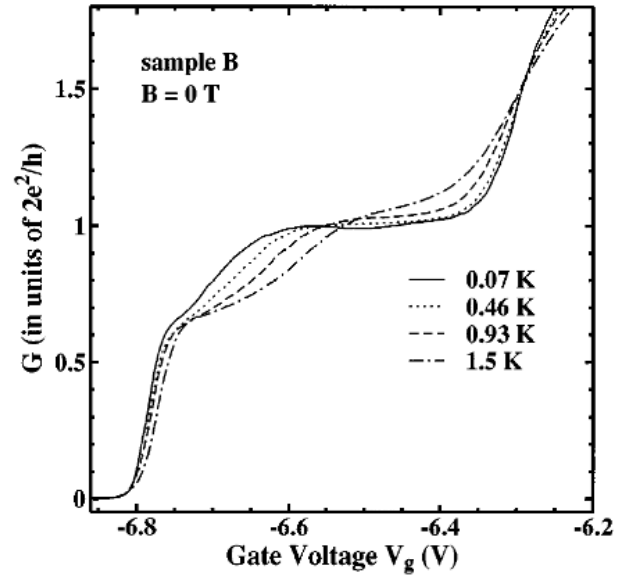}
\caption{Hegðun skotleiðninnar í skammtapuntksnertu sem fall af hliðspennunni $V_g$ fyrir $B=0$ T og nokkur mismunandi hitastig. Niðurstöður teknar úr \cite{Pepper07}.}
\label{FigPepper07}
\end{figure}

Hins vegar er hægt að sjá frávik frá \eqref{LBformula} í mælingum á skotleiðni ef rafeindaþéttleikinn er lítill og einungis eitt leiðniástand er opið. Nánar tiltekið birtist oft dularfullt aukaþrep í leiðninni í kringum $G \approx 0.7G_0$ \cite{Pepper07,Pepper07-1,Pepper07-2}. Þetta má sjá á grafi á mynd \ref{FigPepper07}.

\begin{figure}[tbp]
\includegraphics[width=1.0\columnwidth,keepaspectratio]{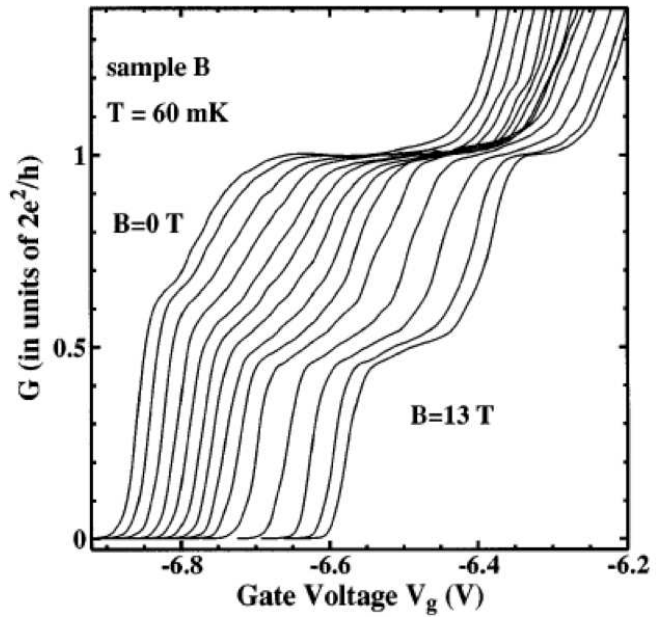}
\caption{Hegðun skotleiðninnar í skammtapuntksnertu sem fall af hliðspennunni $V_g$ fyrir $T=0.60$ mK og mismunandi segulsviðsstyrk $B$. Niðurstöður teknar úr \cite{Pepper07}.}
\label{FigPepper07B}
\end{figure}

Þótt þessu fráviki svipi formlega til brota-skammtahrifa Halls \cite{FQHE}, þá eru eðlisfræðileg upptök \LDQ0.7 fráviksins\RDQ að sjálfsögðu mjög frábrugðin. Tvær staðreyndir úr tilraunum benda til þess að þetta sé tengt spuna á einhvern hátt. Í fyrsta lagi kom í ljós að við það að minnka rafeindaþéttleikann eykst rafeinda g-stuðullinn svo um munar \cite{Pepper07}. Í öðru lagi færist aukaþrepið samfellt frá $0.7G_0$ til $0.5G_0$ ef beitt er segulsviði á sýnið og það aukið smám saman (sjá mynd \ref{FigPepper07B}). Leiðniþrepið við $G=0.5G_0$ í sterku segulsviði er vel skilið út frá Zeeman hrifum.

Nýlegri tilraunaniðurstöður sýna að brotaskömmtun skotleiðninnar er ekki endilega altæk. Svo virðist sem að í löngum skammtavírum birtist aukaþrepið nær $0.5G_0$ en $0.7G_0$ \cite{Reilly}. Einnig er líklegt að munur sé á sýnum með rafeindaleiðni og þeim með holuleiðni. Í seinna tilvikinu eru tilraunaniðurstöður ansi óræðar. Fyrstu niðurstöður fyrir holukerfi bentu til þess að hegðunin sé breytileg milli sýna og sé mjög háð orkubilinu milli léttra og þungra hola \cite{BagraevSemiconductors}. Aðrir hópar sögðust sjá aukaþrepið á sama stað og fyrir rafeindakerfi \cite{Rokhinson,Hamilton}. Staðan er því langt frá því að vera augljós.

Í þessari yfirferð fjöllum við um fræðina bakvið fjölda fyrirbrigða \LDQ0.7 frávikinu\RDQ, sér í lagi í einvíðum kerfum með annað hvort n- eða p-leiðni. Í kafla II verður farið yfir fjölda fræðilíkana. Í þriðja kafla kynnum við hugmynd okkar að skýringu á brotaskömmtun skotleiðninnar sem byggir á Heisenberg skiptaverkun milli staðbundinna og frjálsra rafeinda. Í kafla IV er tekið fyrir tilfellið með holuleiðni og sýnt að hún getur verið ansi frábrugðin rafeindaleiðni. Í niðurlagi verða dregnar saman megin niðurstöðurnar.

\section{Fræðiyfirsýn}

Frá því \LDQ0.7-frávikið\RDQ sást fyrst í tilraunum hefur það reynst eðlisfræðingum á sviði fræðilegrar þéttefniseðlisfræði mjög erfitt að útskýra. Tillögur hafa verið allt frá Wigner-grindar myndun cite{key-1} og skortur á Fermi-vökva hegðun í einvíðum rafgösum \cite
{key-2}, að rafeinda-hljóðeinda víxlverkun \cite{key-3}. Hver þessara ústkýringa hefur náð að útskýra einhverja hlið á fyrirbrigðinu en enn þá vantar kenningu sem er sjálfri sér samkvæm í að útskýra allar hliðar fyrirbrigðisins.

Í þessum kafla förum við yfir þær hugmyndir sem okkur virðast skynsamlegastar og tengjast hugmynd okkar um að brotaskömmtunin leiði af skiptaverkum milli staðbundins spuna og leiðnirafbera \cite{key-4}.

Við byrjum á fenómenólógísku líkani sem útskýrir aukaþrepið með myndun bils milli spuna-orkuborðanna í skammtapunktinum. Hugmyndin, sem upprunnin er hjá Bruus et al. \cite{key-34}, virtist snemma mjög lofandi og var hægt að leiða út einfalda lokaða framsetningu á leiðninni sem fall af efnamættinu og hitastigi. Líkanið hjá Reilly et al. \cite{key-10,key-11,key-25} er flóknasta útgáfan af hugmyndinni. Reilly gerir ráð fyrir að spunabilið sé ekki fasti, heldur sé línulega háð hliðspennunni $V_{g}$. Líkanið tekur inn tvær stýribreytur. Annars vegar er það rýmd punktsnertunnar, sem stýrir því hvernig Fermi orka einvíða rafgassins breytist með $V_g$, og hins vegar er það fastinn $\gamma=d\Delta E_{\uparrow\downarrow} /dV_{g}$, þ.e. hversu hratt spunabilið breikkar sem fall af $V_{g}$. Mismunandi gildi stýribreytanna geta sýnt fram á nokkur stig 0.7-frábrigðisins. Við lágt hitastig gefur líkanið aukaþrep við $0.5 G_0$ sem er tengt við fullkomlega spunaskautaða leiðni. Hið venjulega leiðnigildi $G_0$ fæst þegar hliðarspennan er stillt þannig að efri spuna-orkuborðinn nær niður fyrir Fermi-orku tvívíða rafgassins. Ef hitastigið er hækkað eykst gildi aukaþrepsins vegna örvaða rafeinda frá spuna-orkuborðanum sem tekur ekki þátt í leiðninni, en straumurinn helst nánast stöðugur á tilteknu bili $V_g$ vegna þess að borðabilið eykst jafnt og þétt. Þegar borðabilið $\Delta E_{\uparrow\downarrow} $ er miklu minna en varmaörvunin $k_B T$ myndast hið umtalaða aukaþrep.

Einn höfunda komst að svipaðri niðurstöðu með fenómenólógísku líkani af skammtavír með hálfskautuðu rafgasi. Ástandssætnin var nálguð með þrepafallinu
$n(q,T)=[1+f(T,\varepsilon(q)-\mu+\Delta E_{\uparrow\downarrow})]f(T,\varepsilon(q)-\mu)$
þar sem $f$ er Fermi fallið og efnamættið $\mu$ ræðst af rafberaþéttleikanum $n_{1D}$. Ef gert er ráð fyrir að á ákveðnu bili  $n_{1D}$ sé $\mu(n_{1D})-\Delta E_{\uparrow\downarrow}(n_{1D}) =-\chi = $fasti, og ef skoðað er frekar lágt hitastig, þá fæst aukaþrep í leiðninni við:

\begin{equation}
G=\frac{e^{2}}{h}\left\{ 1+\frac{1}{\exp\left(\xi/kT\right)+1}\right\} .
\label{eq:Shelykh_01}
\end{equation}

Ef hitastig er aukið færist þrepið frá $(1/2)G_0$ í $(3/4)G_0$, sem svarar nokkurn veginn til 0.7 eiginleikans.

Uppruna spunabilsins má rekja til sjálfkrafa spunaskautun vegna skiptaverkunar í tilfelli lítils rafberaþéttleika. Hægt er að útskýra hrifin á eftirfarandi hátt. Línulegur þéttleiki hreyfiorkunnar í einvíða rafgasinu er í réttu hlutfalli við þriðja veldi línulegs þéttleika eindanna, $n_{1D}^3$:

\begin{equation}
\varepsilon _{kin}^{1D}=\frac{\pi ^{2}\hbar ^{2}n_{1D}^{3}}{6mg_{s}^{2}},
\label{eq:Bagraev_03}
\end{equation}
þar sem $g_s$ er spunamargfaldarinn, þ.e. fjöldi rafeinda á einingarrúmmál í fasarúminu. Orkuþéttleikinn er auðvitað í lágmarki fyrir óskautað rafgas með $g_s=2$. Hins vegar er hægt að áætla fyrir skiptaverkunarorkuna að $\epsilon_{sk} \thicksim -n_{1D}^2 /g_s$ og lækkar því við aukna skautun rafgassins. Samspil þessara liða veldur sjálfkrafa skautun við lítinn þéttleika þegar annað veldið sigrar þriðja, en rafgasið verður óskautað í hinu tilfellinu, mikinn þéttleika. Þetta svarar til hins þekkta samseglandi óstöðugleika Stoners, sem var nýlega sýndur í tilraunum með tvívíð kerfi \cite{key-35}.

Með Hartree-Fock nálgun er hægt að áætla skiptaverkunarorkuþéttleikann sem \cite{key-13}

\begin{eqnarray}
\epsilon_{exc}=-\frac{1}{2L}\sum_{E_{K},E_{K}<E_{F}}\left\langle KL\right|
V\left| LK\right\rangle  \nonumber \\
\,\,\,\,\,\,\,\,\,\,\,\approx\frac{0.28e^{2}}{g_{s}}n_{1D}^{2}+\frac{e^{2}}{%
4g_{s}}n_{1D}^{2}\ln \left( \frac{n_{1D}R}{\pi g_{s}}\right),
\label{eq:Bagraev_02}
\end{eqnarray}
þar sem $R$ og $L$ eru radíus vírsins og lengd hans. Ásamt jöfnu \eqref{eq:Bagraev_03} er þá hægt að áætla spunabilið sem

\begin{equation}
\Delta E_{\uparrow\downarrow}\approx2n_{1D}e^2\left[0.15-0.25\ln \left(
\frac{n_{1D}R}{\pi}\right)\right]-\frac{\pi ^{2}\hbar ^{2}n_{1D}^{2}}{2m}.
\label{SpinGap}
\end{equation}
Línulega þéttleikann má áætla sem fall af hliðspennunni og rýmdinni $c$ á milli punktsnertunnar og einvíðu rafskautanna, $n_{1D}=cV_{sg}/e$. Fyrri liðurinn í \eqref{SpinGap} lýsir samfelldri opnun bilsins þegar við erum með lítinn þéttleika og skiptaverkunin er ráðandi. Hann er innifalinn í líkani Reilly og þar með er hægt að meta fenómenólógísku breytuna hans $\gamma$. Hins vegar tekur hann ekki með seinni liðinn þar sem hreyfiorkan veldur samfalli bilsins.

Wang og Berggren skoðuðu einnig spunabil vegna skiptaverkunar í óendanlegum hálf-einvíðum vírum með því að nota tölulega DFT (e. density functional theory)  \cite{key-27}. Þeir fengu svipaða niðurstöðu og að ofan, þ.e. stórt orkuborðabil og fullkomna skautun við lítinn þéttleika. Einnig var lögð áhersla á að þessi hrif má bæla að hluta til með því að taka inn í myndina fylgni áhrif (það sama á við Hartree-Fock aðferðina að ofan).

Öll þessi fenómenólógísku líkön hafa tvo galla. Í fyrsta lagi spá þau fyrir um það að gildi aukaþrepsins sé ekki altækt og geti verið einhvers staðar á bilinu $0.5 G_0$ til $G_0$ eftir því hvernig stýribreytur eru valdar. Í öðru lagi lækkar gildi aukaþrepsins með lækkandi hitastigi og nær $0.5 G_0$ þegar $T\rightarrow 0 $, en það brýtur í bága við tilraunaniðurstöður.

Til að reyna að fanga hitastigstengslin sem sjást í tilraunum, komu Yigal Meir og samstarfsmenn með tillögu sem byggir á fjöleinda Kondo fræði \cite{key-8}. Þessi hugmynd fær stuðning frá því að hópur frá Harvard benti á ákveðin atriði sem eru svipuð með Kondo-leiðninni í skammtapunktum og leiðninni í punktsnertum nálægt 0.7-frábrigðinu \cite{key-16}, nefnilega:

i) Hvass leiðnitoppur myndast við Fermi orkuna við lágt hitastig (núll-spennu frábrigði)

ii) Leiðnina er altækt hægt að stilla með einni skölunarbreytu fyrir breitt svið hliðarspennu og skölunarbreytuna má kalla Kondo hitastig kerfisins, $T_K$,

\begin{equation}
G=2e^{2}/h\left[1/2f\left(T/T_{K}\right)+1/2\right].  \label{eq:Meir_01}
\end{equation}
Altæka fallið $f(T/T_K)$ er hið sama fyrir skammtapunkta nálægt Kondo hrifum og er í rauninni eini munurinn á kerfunum margfaldararnir (1/2) í \eqref{eq:Meir_01}. Við höfum hliðarskilyrðin $f(0)=1$ og $f(\infty)=0$.

iii) Núll-spennu toppurinn verður fyrir Zeeman hrifum í sterku segulsviði.

Atriði ii), sem gefur altækt leiðnigildi við hátt hitastig, $0.5G_0$, veitti Yigal Meir og samstarfsmönnum innblástur til að lýsa skammtapunktsnertu nálægt 0.7 frábrigðinu með eftirfarandi Hamiltonvirkja:

\begin{align}
H&= \sum_{k \sigma \in L,R} \epsilon_k c_{k\sigma}^\dagger c_{k\sigma} +\sum_\sigma \epsilon_{d\sigma} d_\sigma^\dagger d_\sigma + U n_{d\uparrow}n_{d\downarrow} \nonumber\\
&+ \sum_{k\sigma} \left[ v_k^{(1)} \left( 1-n_{d \overline \sigma } \right) c_{k\sigma}^\dagger d_\sigma + V_k^{(2)}n_{d \overline \sigma} c_{k\sigma} ^\dagger d_\sigma +h.c. \right]
\label{eq:Meir_02}
\end{align}
þar sem $c_{k\sigma}$ er eyðingarvirki rafeindar í skauti með spuna $\sigma$ og $d_{\sigma}$ eyðingarvirki rafeindar sem er staðbundin í punktsnertunni. Litið var á tengifastana $V_k^{(2)}$, $V_k^{(1)}$ sem orku-óháð þrepaföll til að lýsa hleðslusveiflunum $0\leftrightarrow1$ og $1\leftrightarrow2$ í punksnertunni. Munurinn á þessum virkja og upphaflega Anderson Hamiltonvirkjanum er sá að $V_k^{(2)} <V_k^{(1)}$ sem tengist því að ef punktsnertan er með staka rafeind þá minnkar samtvinnun punktsnertunnar og skautanna vegna Coulomb fráhrindikraftanna, ef miðað er við tóma punktsnertu. Bent var á að við hátt hitastig ætti þessi Hamiltonvirki að gefa leiðnina $G=0.5G_0$ vegna $0\leftrightarrow 1$ sveiflna, en lækkun hitastigs leiðir til Kondo-hækkun leiðninnar vegna $1\leftrightarrow 2$ sveiflna þar til venjulega gildið $2e^2/h$ fæst við $T=0$.

Til að fá ákveðna innsýn var Schrieffer-Wolff umbreytingu beitt á Hamilton-virkjann \cite{key-18} til að varpa honum á Kondo-virkja þar sem skiptatengslin $J_K$ eru háð föstunum $V_k^{(1,2)}$. Þarnæst var farið í truflanareikninga út frá skiptatengslunum til að fá út núll-spennu frábrigðið. Hins vegar fæst með þessari aðferð ekki hið rétta standard gildi $G_0$ skotleiðninnar þegar skautspennan $V_{ds} \rightarrow 0$. Við teljum að hægt sé að skoða þetta tilvik betur með Kubo línulegri fræði þar sem tengifastarnir eru teknir með í reikninginn á nákvæman hátt. Önnur veik hlið ofangreindrar aðferðar er sú að hæð aukaþrepsins er ekki altæk og fer mjög mikið eftir stýribreytum líkansins og hitastigi. Líkanið nær ekki að útskýra af hverju mjög stuttar punktsnertur eru með aukaþrepið mjög stöðugt í hæðinni $0.7G_0$ en ekki t.d. $0.85G_0$.

Einnig skal benda á það að nýlegar tilraunaniðurstöður virðast fara þvert á niðurstöður úr Kondo líkani. Graham og samhöfundar hans segjast í nýlegri grein \cite{key-19} ekki sjá núll-spennu frábrigðið, í það minnsta í sumum tilraunauppsetningum. Önnur tilraun með skammtapunktum sýnir frábrigðið mjög stöðugt allt niður í mjög lágt hitastig \cite{key-33}, á meðan líkanið að ofan spáir því að það hverfi. Hlutverk Kondo fylgni í 0.7 frábriðginu er því enn mjög óljóst.

Útskýring á því að leiðnin er mjög stöðug kringum $0.7G_0$ fæst í líkönum sem byggja á samspili einstigs og þrístigs leiðnirása fyrir rafeindapör. Fyrsta líkanið af þessu tagi kom frá Flambaum og Kuchiev \cite{Flambaum}, en þeir skoðuðu flutning bundinna rafeindapara um skammtapunktsnertur. Ef skiptaverkun gefur einstiginu og þrístiginu mismunandi orkur verður virka mættið sem rafeindapör sjá við að fara gegnum punktsnertuna spunaháð. Sé orka þrístigsins lægri þá er fyrir ákveðið bil á $V_g$ opið fyrir flutning þrístigs para en lokað fyrir einstig. Án segulsviðs eru líkurnar á myndun þrístigs $3/4$ en einstigs $1/4$, þannig að leiðnin ætti þá að vera $(3/4)G_0$. Sömuleiðis ætti að fást leiðnin $(1/4)G_0$ ef orka einstigsins er lægri. Ef leyfð er myndun bundinna ástanda með fleiri en tveimur rafeindum fást fleiri aukaþrep.

Veiki hlekkurinn í líkani Flambaum og Kuchiev er ímyndað aðdráttarafl milli rafeinda til að þær myndi pör. Þótt ekki sé hægt að hafna því fyrirfram að slíkt aðdráttarafl sé til staðar þá er uppruni þess ekki ljós og þess vegna er þessi hugmynd frekar varasöm. Hins vegar er til útgáfa af líkaninu sem byggir á samspili einstigs og þrístigs tvístrunar rása en gerir ekki ráð fyrir neinu aðdráttarafli. Hugmyndin var sett fram af Rejec et al. \cite{key-5,key-6,key-7} og gerir ráð fyrir að bundið ástand einnar rafeindar myndist í punktsnertunni \cite{Bird}, og rafeindin víxlverki við leiðnirafeindir með Heisenberg skiptaverkun. Ef orka leiðnirafeindanna er nógu lítil verður alltaf rafeind eftir bundin að tvístruninni lokinni \cite{key-14,key-15} og hægt er að reikna út flutningslíkurnar með Landauer-Buttiker jöfnunni fyrir alkul:

\begin{equation}
G_{T=0}(E_{F})=\frac{2e^{2}}{h}\left[ \frac{1}{4}\mathrm{T_{s}\left(
E_{F}\right) }+\frac{3}{4}\mathrm{T_{t}\left( E_{F}\right) }\right] ,
\label{eq:rejec_03}
\end{equation}
þar em $T_s$ og $T_t$ eru flutningsfastarnir fyrir einstig og þrístig og stuðlarnir $1/4$ og $3/4$ standa fyrir líkurnar á stigunum. Við hitastig $T\neq 0$ er hægt að útfæra fyrri niðurstöðu:

\begin{equation}
G\left( T,\mu \right) =\int_{0}^{\infty }G_{T=0}\left( E\right) \left( -%
\frac{\partial f\left( T,E,\mu \right) }{\partial E}\right) dE.  \label{EqT}
\end{equation}
Rejec et al. \cite{key-7} héldu því fram að virka mættið sem leiðnirafeindir sjá sé tveir þröskuldar með lágmark í miðju bungunnar. Á milli þeirra myndast hermuástönd sem svara til einstigs og þrístigs og orkumunurinn stafar af skiptaverkuninni. Hermandi smug gegnum þessi ástönd svara undir ákveðnum aðstæðum til tveggja leiðnihásléttna með $G \thicksim (1/4)G_0$ og $G \thicksim(3/4)G_0$. Hins vegar, ef ferkantaðir mættisþröskuldar eru notaðir sést einungis ein slétta, sem fer eftir formerki skiptaverkunarstuðulsins \cite{Flambaum,ShelykhJPCM}.

Fjölrafeinda fylgnin, sem svipar til Kondo fræða \cite{key-21}, valda hitastigsháðri endurnormun skiptaverkunarstuðulsins. Samkvæmt skölunaraðferð Andersons \cite{Anderson} fæst að fyrir andseglandi víxlverkun vex stuðullin mjög við lágt hitastig og myndast því Kondo ský umhverfis staðbundna spunann. Fyrir samseglandi víxlverkun stefnir stuðullinn hins vegar á núll þegar hitastigið gerir það og því hverfur munurinn á einstigi og þrístigi og þar með 0.7 frábrigðið við lág hitastig \cite{ShelykhJPCM}.

Mikið hefur verið rannsakað fræðilega hvort mögulegt sé að í skammtapunktsnertu myndist bundið ástand staks spuna. Niðurstöður úr tölulegri DFT styðja tilgátuna um að annað hvort myndist skautað svæði í punktsnertunni vegna víxverkunar milli rafeinda \cite{key-9,key-36,key-37,key-38} eða þriggja rafeinda andseglandi spunagrind \cite{key-8,key-20} sem virkar þá eins og spuni $1/2$. Fyrra tilvikið getur gegnt hlutverki virkrar spunasíu.

Taka skal fram að til eru tilraunaniðurstöður sem styðja einstigs og þrístigs rásaleiðnina sem lýst er hér og eru úr mælingum á straumsuði í skammtavírum og skammtapunktum \cite{key-29,key-30}. Sér í lagi eru mælingar á Fano stuðlinum, sem sýna frávik frá Poisson dreifðu suði, í samræmi við hlutfallið 3:1 sem þrístigs-einstigs líkindahlutfallið  \cite{key-32}.

Við teljum að þessar niðurstöður geri okkur kleift að líta svo á að 0.7 fyrirbrigðið stafi af skiptaverkun milli leiðnirafbera og staðbundins spuna $J$ í punktsnertunni. Frumvinnu Rejec et al um staka bundna rafeind með $J=1/2$ má útfæra í tilfellið með $N$ staðbundnar rafeindir eða holur. Í því tilfelli verður hæð aukaþrepsins almennt brot margfaldað við $G_0$ en ekki endilega $(3/4)G_0$ eins og í \cite{Flambaum,key-6}. Þetta var hvatinn að kenningu okkar varðandi brotaskömmtun skotleiðninnar sem við munum rekja í eftirfarandi köflum.

\section{Brotaskömmtun skotleiðninnar í einvíðum rafeindakerfum}

Í þessum kafla kynnum við hugmynd okkar um brotaskömmtun skotleiðninnar í einvíðum rafeindakerfum og gefum alhæfingu á líkani Rejec et al fyrir tilfelli þar sem spuni $J>1/2$ er staðbundinn í skammtapunktsnertunni \cite{key-4}. Raunveruleg framkvæmd slíks tilfellis gæti verið t.d. punktnerta með innfelldri Mn$^{2+}$ jón með spuna $J=5/2$. Önnur leið væri punktsnerta með nokkrum staðbundnum rafeindum og við setjum kerfið í ástand þar sem sjálfkrafa skautun á sér stað. Seinna tilfellið á líklega við löngu skammtavírana sem gerðar voru tilraunir á í verki Reilly et al \cite{Reilly}. Eins og þegar hefur verið tekið fram myndast þá aukaþrepið kringum $G \simeq 0.5G_0$ en ekki $G \simeq 0.7G_0$.

Lítum fyrst á þetta á eigindlegan hátt. Með segulspuna $J\neq 0$ hefur bundna ástandið áhrif á leiðnirafberana með Heisenberg skiptaverkun. Leiðnistuðullinn um punktsnertuna virðist því vera spunaháður. Eftir að að leiðnirafeind smýgur inn í punktsnertuna getur heildarspuninn verið annað hvort $S_1=J+1/2$ eða $S_2=J-1/2$. Fjöldi mögulegra uppraðana í hvoru tilfelli er $N_1=2S_1+1=2J+2$ og $N_2=2S_2+1=2J$. Fyrir samseglandi víxlverkun er ástand 1 orkulægra en ástand 2 og því virki mættisþröskuldurinn sem myndast í punktsnertunni hærri fyrir ástand 2. Þar af leiðandi gerist það ef efnamættið er nógu lítið, að flutningsrafeind í ástandi 1 fer greiðlega gegnum punktsnertuna en rafeind í ástandi 2 tvístrast tilbaka. Þar af leiðandi taka einungis rafeindir í ástandi 1 þátt í leiðninni. Án segulsviðs eru líkurnar á ástandi 1 $(J+1)/(2J+1)$ þannig að leiðnin verður

\begin{equation}
G_{f}=\frac{J+1}{2J+1}G_{0}.  \label{Eq1}
\end{equation}
Til samanburðar fæst ef víxlverkunin er andseglandi að ástand 2 er orkulægra og leiðnin verður

\begin{equation}
G_{a}=\frac{J}{2J+1}G_{0}.  \label{Eq2}
\end{equation}
Tafla 1 gefur samantekt af mögulegum leiðnigildum fyrir mismunandi $J$. Fyrir samseglandi víxlverkun, sem er raunhæfasta tilfellið í tilraunum, sést að gildi aukaþrepsins lækkar með fjölda bundinna rafeinda og nær gildinu $0.5G_0$ þegar $J\rightarrow \infty$. Fjöldi óparaðra rafeinda fer væntanlega eftir lengd punktsnertunnar þannig að fyrir stutta punktsnertu má vænta leiðni kringum $0.7G_0$ á meðan í löngum vírum verður hún $0.5G_0$. Þetta kemur algjörlega heim og saman við niðurstöður tilrauna Reilly et al \cite{Reilly,key-10,key-11}. Ef kveikt er á segulsviði skautast bæði leiðni- og bundnar rafeindir þannig að leiðnin verður $0.5G_0$ fyrir öll $J$ eins og sést í tilraunum.

\begin{table}[tbp]
\caption{Möguleg leiðnigildi fyrir mismunandi $J$.}
\begin{tabular}{|l|l|l|} \hline
Gildi á $J$ & $G_{s}$, samseglandi & $G_{a}$, andseglandi \\\hline
$J=1/2$ & $G_{f}=3/4G_{0}$ & $G_{a}=1/2G_{0}$ \\
$J=1$ & $G_{f}=2/3G_{0}$ & $G_{a}=1/3G_{0}$ \\
$J=3/2$ & $G_{f}=5/8G_{0}$ & $G_{a}=3/8G_{0}$ \\
$J=2$ & $G_{f}=3/5G_{0}$ & $G_{a}=2/5G_{0}$ \\
$J=5/2$ & $G_{f}=7/12G_{0}$ & $G_{a}=5/12G_{0}$ \\
$J=\infty$  & $G_{f}=1/2G_{0}$ & $G_{a}=1/2G_{0}$ \\ \hline
\end{tabular}
\end{table}

Taka skal fram að gagnstætt við \cite{key-7} þá er í okkar líkani ekki hægt að sjá tvö þrep samtímis (t.d. $0.75G_0$ og $0.25G_0$). Munurinn felst í hvers konar virkt mætti rafeindirnar rekast á í punktsnertunni. Hjá okkur er einungis einn þröskuldur með spuna-háða hæð, en hjá \cite{key-7} eru tveir þröskuldar með hermu-ástönd á milli sem svara til einstigs og þrístigs sem eru klofin vegna skiptaverkunar.

Nánari útreikningar á leiðninni byggja á að reikna út vísa fyrir mismunandi flutningsferli gegnum punktsnertuna, þ.e. flutning án spunaskipta og flutning þar sem spunaskammtur flyst á milli leiðnirafeindar og staðbundinnar. Þessir vísar eru háðir hliðspennunni og leiðnin er svo í réttu hlutfalli við summu annara velda allra vísanna. Farið er nákvæmlega í þessa útreikninga í \cite{english} en við sýnum hér einungis niðurstöðurnar.

\begin{figure*}[!htbp]
\includegraphics[width=2.0\columnwidth,keepaspectratio]{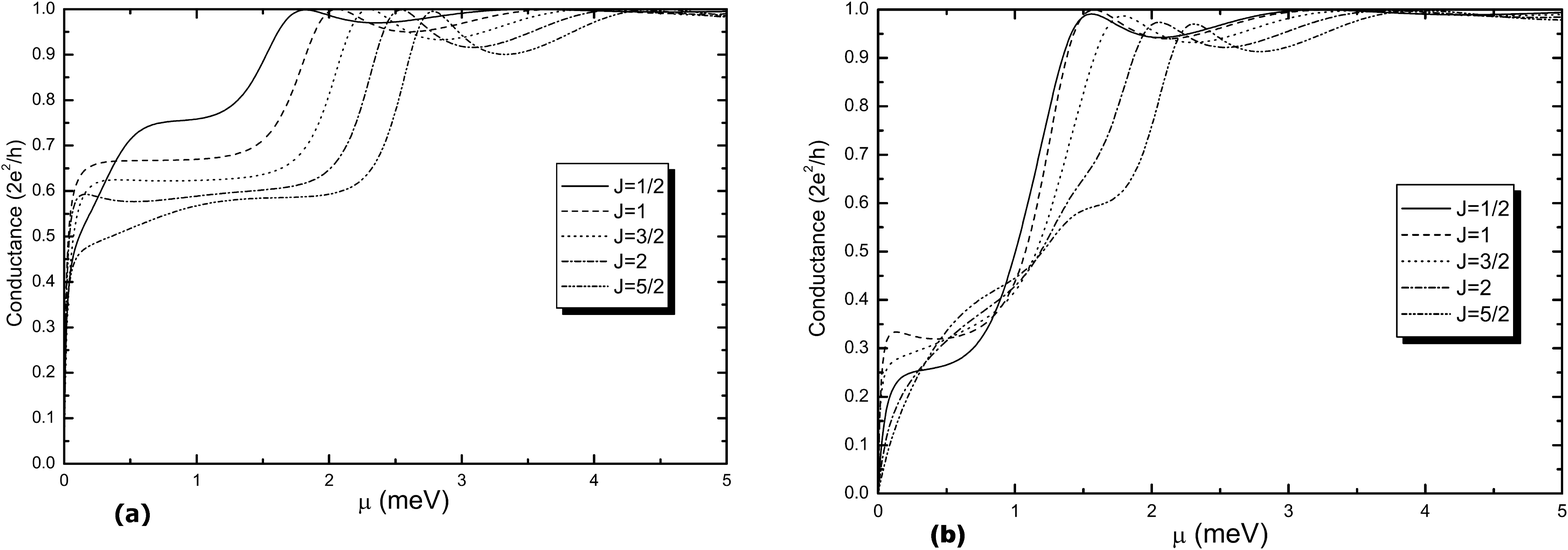}
\caption{Hegðun leiðninnar í skammtapunktsnertunni sem fall af efnamættinu fyrir mismunandi gildi á staðbundna spunanum $J$. Stýribreyturnar í útreikningunum voru eftirfarandi: virkur massi rafbera $m_v=0.06m_e$, hitastigið $T=4$ K, lengd snertunnar $L=L_0 J$ með $L_0=50$ nm, $V_b= 1$ meV, $|V_{sk}|=0.48$ meV fyrir (a) samseglandi og (b) andseglandi víxlverkun.}
\label{FigM1}
\end{figure*}

Mynd \ref{FigM1} sýnir hvernig leiðni skammtapunktsnertunnar breytist með efnamættinu fyrir mismunandi gildi á staðbundna spunanum $J$.  Greinilegt er að fyrir $J>1/2$ fást aukaþrep í skotleiðninni með gildi frábrugðin $0.75G_0$ en það kemur heim og saman við jöfnur \eqref{Eq1} og \eqref{Eq2}.

Svipað og var sagt um staka staðbundna rafeind í kafla 2 ætti fjöl-rafeinda fylgni að valda hitastigsháðri endurnormun stuðulsins $V_{sk}$ og eyðingu aukaþrepsins við lág hitastig.

\section{Brotaskömmtun skotleiðninnar í einvíðum holukerfum}

Eins og þegar hefur komið fram getur brotaskömmtun skotleiðninnar verið eigindlega mismunandi eftir því hvort rafberarnir í kerfinu eru rafeindir eða holur. Ástæðan er mismunandi spunagerð rafeinda og hola í hálfleiðurum eins og Si, Ge og GaAs. Gildisborðar þessara hálfleiðara samanstanda af borða þungra hola með spuna $JJ_z^{þh}=\pm 3/2$, og borða léttra hola með spuna $JJ_z^{lh}=\pm 1/2$ \cite{SplitOFF}. Sökum skömmtunarhrifa myndast orkubil $\Delta$ milli þessara borða í einni og tveimur víddum. Stærð $\Delta$ fer eftir breidd skammtavírsins, muninum á virkum massa léttu og þungu holanna, $m_{lh}$ og $m_{þh}$, og álagi.

Spunaháða tvístrun staðbundinna og leiðni hola má skoða á svipaðan hátt og rafeindirnar í fyrri kafla. Við athugum flutning frjálsra hola gegnum virkan mættisþröskuld tilkominn vegna spunaháðrar víxlverkunar, $V_{b}+V_{sk} \textbf{J}_l\cdot \textbf{J}_{st}$ þar sem við gerum ráð fyrir að hola sé staðbundin í punktsnertunni. Hér á $l$ við leiðniholur, $st$ við staðbundnar, $V_b>0$ er bein Coulomb víxlverkun og $V_{sk}$ er skiptaverkunarstuðullinn.

Við skulum fyrst reyna að skilja eigindlega hvernig munurinn á spunagerð rafeinda og hola endurspeglast í munstrinu á brotaskömmtun skotleiðninnar. Hægt er að endurskrifa spunaháða hluta virka Hamiltonvirkjans fyrir holur sem $V_{sk} \textbf{J}_l \cdot \textbf{J}_{st} = V_{sk} ( \textbf{J}_T^2- \textbf{J}_l^2- \textbf{J}_{st}^2)/2 = V_{ex} [J_T(J_T+1)-3(3/2+1)]/2$. Þar af leiðandi fást fyrir mögulegu algildi heildarspuna holuparsins, $J_T=3,2,1,0$, eftirfarandi gildi á hæð mættisþröskuldsins: $V_{b}+(9/4)V_{sk}$, $V_{b}-(3/4)V_{sk}$, $V_{b}-(11/4)V_{sk}$, $V_{b}-(15/4)V_{sk}$. Fyrir andseglandi víxlverkun er þröskuldurinn lægstur ef hæsta mögulega spuna er náð, en í samseglandi tilvikinu er það öfugt. Ef við höfum óskautað upphafsástand ($B=0$) eru líkurnar á að fá heildarspuna $J_T=3,2,1,0$ eftirfarandi: $7/16$, $5/16$, $3/16$ og $1/16$. Þar af leiðandi, ef við skoðum ímyndaða tilvikið að $\Delta=0$ \cite{note} má gera ráð fyrir að við fáum leiðniþrep nálægt gildunum $(7/16)G_0$, $(12/16)G_0$ og $(15/16)G_0$ (ef leiðnin í þessu kerfi er núna skömmtuð með $G_0=4e^2/h$ þar sem tveir borðar taka þátt í leiðninni). Í andseglandi tilvikinu fást í staðinn gildin $e^2/4h$, $e^2/h$ og $9e^2/4h$.

Niðurstöðurnar geta verið annars eðlis ef bilið milli þungra og léttra hola er ekki mjög lítið. Til dæmis ef $\Delta \rightarrow \infty$, þá er einungis þungu holu borðinn í boði fyrir bæði staðbundnar og leiðniholur. Leiðnin í kerfi án víxlverkunar skammtast því í einingunni $G_0=2e^2/h$ eins og fyrir rafeindir. Spunar pöruðu holanna geta verið annað hvort samsíða eða andsamsíða með jafn háum líkum. Vegna þess hve léttu holu borðinn er fjarlægur í orku eru spunasveiflur ekki leyfðar \cite{SpinFlipHoles} og þess má vænta að sjá einungis eitt aukaþrep með $G=e^2/h$. Þetta er því mjög frábrugðið því þegar um rafeindir er að ræða, en þá eru spunasveiflur leyfðar og þar af leiðandi fæst aukaþrep með $G=3e^2/h$ \cite{Flambaum,key-5,key-7}.

Í rauninni er $\Delta$ einhver endanleg stærð og verður málið því flóknara en hér að ofan. Gera þarf nákvæmari greiningu á líkaninu í þessu tilviki en hana má sjá í \cite{english}. Hér rekjum við einungis niðurstöðurnar.

\begin{figure*}[tbp]
\includegraphics[width=2.0\columnwidth,keepaspectratio]{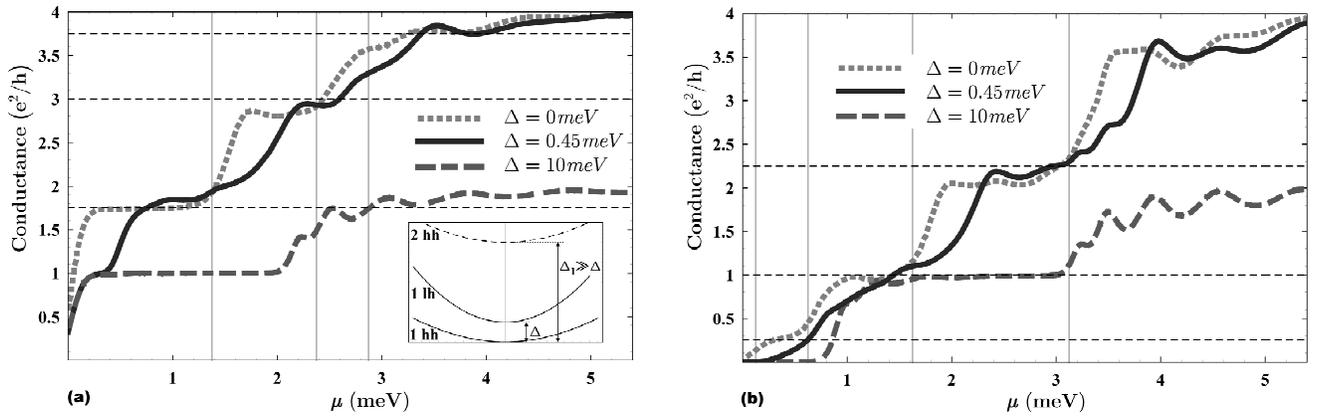}
\caption{(a) Þrepin í leiðnitröppuni sem fall af efnamætti leiðnihola í GaAs skammtapunktsnertu fyrir samseglandi víxlverkun. Ferlarnir svara til þriggja gilda á $\Delta$, $\Delta (meV)=0,0.45,10$. Stuðlarnir fyrir beina Coulomb víxlverkun og skiptaverkun voru áætlaðir sem $V_{b}\simeq 1meV$ og $V_{sk}\simeq -0.5meV$. Lóðréttu gráu línurnar svara til hæðar virka mættisþröskuldsins, $V_{b}-3V_{sk}/4$, $V_{b}-11V_{sk}/4$ and $V_{b}-15V_{sk}/4$ á meðan láréttu línurnar svara til gildanna $7e^{2}/4h$, $3e^{2}/h$ og $15e^{2}/4h$. Litla myndin sýnir borðagerð einvíðu holanna. (b) Sama en fyrir andseglandi víxlverkun, $V_{sk}\simeq 0.5meV$ og $V_{b}\simeq 2meV$. Láréttu línurnar svara til gildanna $e^{2}/4h$, $e^{2}/h$ and $9e^{2}/4h$. }
\label{FigM3}
\end{figure*}

Myndir \eqref{FigM3} sýna leiðni í GaAs skammtapunktsnertu fyrir mismunandi orkubil $\Delta$. Sé $\Delta$ tiltölulega lítið fást þrepin sem búist var við nálægt $7e^{2}/4h$, $3e^{2}/h$ og $15e^{2}/4h$ í samseglandi tilvikinu en nálægt $e^{2}/4h$, $e^{2}/h$ og $9e^{2}/4h$ ef víxlverkunin er andseglandi. Þar að auki er eitt þrep í $e^2/h$ í samseglandi tilvikinu. Taka skal fram að skömmtun skotleiðninnar í Si og Ge kerfum með holuleiðni er væntanlega svipuð og í GaAs sökum svipaðrar spunagerðar. Hins vegar getur útlitið verið annað í IV-VI hálfleiðurum eins og PbTe, PbSe og PbS þar sem rafeinda-holu samhverfa ríkir.

\section{Niðurlag}
Sýnt var fram á að ef stakur spuni $J$ er staðbundinn í skammtapunktsnertunni má sjá brotaskömmtun skotleiðninnar. Fyrir sýni með rafeindaleiðni valda stakar staðbundnar rafeindir með spuna $J=1/2$ stöku aukaþrepi með leiðni $G=0.75G_0$. Aukning á $J$ sem fylgir lengingu á punksnertunni veldur því að aukaþrepið lækkar niður í $0.5G_0$. Í sýnum með holuleiðni er staðan töluvert flóknari sökum flókinnar spunagerðar efna eins og Si, Ge, og GaAs. Eitt eða fleiri brotaskömmtunarþrep myndast eftir þvi hver orkumunurinn er á þungum og léttum holum.

\section{Þakkir}
Höfundar þakka prófessor N. T. Bagraev fyrir hjálpsemi hans og samræður. Þakkir eru veittar fyrir stuðning Vísinda- og tækniráðuneytis Brasilíu, IBEM, Brasilíu, og FP7 IRSES verkefnisins \LDQ SPINMET\RDQ.


\begin{thebibliography}{99}
\bibitem{Landauer}  R. Landauer, \textit{IBM J. Res. Dev.} \textbf{1}, 233
(1957).

\bibitem{Buttiker}  M. Buttiker, \textit{Phys. Rev. Lett.} \textbf{57},
1761(1986).

\bibitem{Wees}  B.J. van Wees, H. van Houten, C.W.J. Beenakker, J.G.
Williamson, L.P. Kouwenhoven, D. van der Marel, C.T. Foxon, \textit{Phys.
Rev. Lett.} \textbf{60}, 848 (1988).

\bibitem{Wharam}  D.A. Wharam, T.J. Thornton, R. Newbury, M. Pepper, H.
Ahmed, J.E.F. Frost, D.G. Hasko, D.C. Peacok, D.A. Ritchie, G.A.C. Jones,
\textit{et al.}, \textit{J. Phys. C} \textbf{21}, L209 (1988).

\bibitem{Pepper07}  K.J. Thomas, J.T. Nicholls, M.Y. Simmons, M. Pepper,
D.R. Mace, D.A. Ritchie, \emph{Phys. Rev. Lett.} \textbf{77}, 135 (1996).

\bibitem{Pepper07-1}  K.J. Thomas, J.T. Nicholls, N.J. Appleyard, M.Y.
Simmons, M. Pepper, D.R. Mace, W.R. Tribe, D.A.Ritchie., \emph{Phys. Rev. B}
\textbf{58}, 4846 (1998).

\bibitem{Pepper07-2}  K.J. Thomas, J.T. Nicholls, M. Pepper, W.R. Tribe,
M.Y. Simmons, D.A. Ritchie. \emph{Phys. Rev. B }\textbf{61}, R13365 (2000).

\bibitem{FQHE}  H.L. Stormer, A.Chang, D.C. Tsui, H.C.M. Hvang, A.C.
Gossard, W. Wiegman, \emph{Phys. Rev. Lett.} \textbf{50}, 1953 (1983)

\bibitem{Reilly}  D. J. Reilly, G. R. Facer, A. S. Dzurak, B. E. Kane, R. G.
Clark, P. J. Stiles, A. R. Hamilton, J. L. O'brien, N. E. Lumpkin, L. N.
Pfeiffer, and K. W. West, \textit{Phys. Rev. B} \textbf{63}, R121311 (2001)

\bibitem{BagraevSemiconductors}  N.T. Bagraev , A.D. Bouravleuv, L.E.
Klyachkin, A.M. Malyarenko, W.Gehlhoff, V.K.Ivanov and I.A. Shelykh, \emph{%
Semiconductors} \textbf{36}, 439 (2002)

\bibitem{Rokhinson}  L.P. Rokhinson, L.N Pfeiffer, K.W. West, \textit{Phys.
Rev.Lett.} \textbf{96}, 156602 (2006).

\bibitem{Hamilton}  O. Klochan, W. R. Clarke, R. Danneau, A. P. Micolich, L.
H. Ho, A. R. Hamilton, K. Muraki, and Y. Hirayama, \emph{Appl. Phys. Lett.}
\textbf{89}, 092105 (2006).

\bibitem{key-1}  K. A. Matveev, \textit{Phys. Rev. Lett.} \textbf{92},
106801 (2004).

\bibitem{key-2}  L. Bartosch, M. Kollar, and P. Kopietz, \textit{Phys. Rev. B%
}, \textbf{67}, 092403 (2003).

\bibitem{key-3}  G. Seelig and K. A. Mateev, \textit{Phys. Rev. Lett.}
\textbf{90}, 176804 (2003)

\bibitem{key-4}  I. I.A. Shelykh, N.G. Galkin and N.T. Bagraev, \textit{%
Phys. Rev. B} \textbf{74}, 085322 (2006).

\bibitem{key-34}  H. Bruus, V.V. Cheianov, and K. Flensberg, \textit{Physica
E} \textbf{10}, 97 (2001).

\bibitem{key-10}  D. J. Reilly, T. M. Buehler, J. L. O'Brien, A. R.
Hamilton, A. S. Dzurak, R. G. Clark, B. E. Kane, L. N. Pfeiffer, and K. W.
West, \textit{Phys. Rev. Lett.} \textbf{89}, 246801 (2002)

\bibitem{key-11}  D. J. Reilly, \textit{Phys. Rev. B} \textbf{72}, 033309
(2005).

\bibitem{key-25}  D. J. Reilly, \textit{Proc. 2nd Quantum Transport
Nano-Hana International Workshop IAP} Conf. Series 5 pp.7-11 (2004).

\bibitem{key-28}  I.A. Shelykh, N.T. Bagraev, V.K. Ivanov, and L. E.
Klyachkin, \textit{Semiconductors} \textbf{36}, 65 (2002).

\bibitem{key-35}  A. Ghosh, C. J. Ford, M. Pepper, H. E. Beere, and D. A.
Ritchie, \emph{Phys. Rev. Lett.} \textbf{92}, 116601 (2004).

\bibitem{key-13}  N.T. Bagraev, I. A. Shelykh, V. K. Ivanov, and L. E.
Klyachkin, \textit{Phys. Rev. B} \textbf{70}, 155315 (2004).

\bibitem{key-27}  Chuan-Kui Wang and K.F. Berggren, \textit{Phys. Rev. B}
\textbf{54}, 14257 (1996)

\bibitem{key-8}  Y. Meir, K. Hirose and N. S. Wingreen, \textit{Phys.
Rev.Lett.} \textbf{89}, 196802 (2002).

\bibitem{key-16}  S. M. Cronenwett, H. J. Lynch, D. Goldhaber-Gordon, L. P.
Kouwenhoven, C. M. Marcus, K. Hirose, N. S. Wingreen, and V. Umansky,
\textit{Phys. Rev. Lett.} \textbf{88}, 226805 (2002).

\bibitem{key-17}  D. Goldhaber-Gordon, J. Gores, M. A. Kastner, Hadas
Shtrikman, D. Mahalu, and U. Meirav, \textit{Phys. Rev. Lett.} \textbf{81},
5225 (1998).

\bibitem{key-23}  D. Goldhaber-Gordon, Hadas Shtrikman, D. Mahalu, David
Abusch-Magder, U. Meirav, and M. A. Kastner, \textit{Nature} (London)
\textbf{391}, 156 (1998).

\bibitem{key-18}  J. R. Shrieffer and P. A. Wolff, \textit{Phys. Rev.}
\textbf{149}, 491 (1966)

\bibitem{key-19}  A.C. Graham and M. Pepper, \textit{Phys. Rev. B}, \textbf{%
75}, 035331 (2007)

\bibitem{key-33}  Yunchul Chung, Sanghyun Jo, Dong-In Chang, Hu-Jong Lee, M.
Zaffalon, V. Umansky, and M. Heiblum, \textit{Phys. Rev. B}, \textbf{76},
035316 (2007).

\bibitem{Flambaum}  V.V. Flambaum and M.Yu. Kuchiev, \textit{Phys. Rev. B}
\textbf{61} R7869 (2000)

\bibitem{key-5}  T. Rejec and A. Ramsak, \textit{Phys. Rev. B} \textbf{62},
12985 (2000).

\bibitem{key-6}  T. Rejec, A. Ramsak, and J. H. Jefferson, \textit{J. Phys.:
Condens. Matter} \textbf{12}, 233 (2000).

\bibitem{key-7}  T. Rejec, A. Ramsak, and J. H. Jefferson, \textit{Phys.
Rev. B}, \textbf{67} 075311 (2003).

\bibitem{Bird}  Möguleiki á staðbundnum rafberum í skammtapunktsnertum var nýlega rannsakaður í tilraun eftir Y. Yoon,
L. Mourokh, T. Morimoto, N. Aoki, Y. Ochiai, J. L. Reno, and J. P.
Bird, \emph{Phys. Rev. Lett.} \textbf{99}, 136805 (2007)

\bibitem{key-14}  J. R. Oppenheimer, \textit{Phys. Rev.} \textbf{32}, 361
(1928)

\bibitem{key-15}  N. F. Mott, \textit{Proc. R. Soc. London} Ser. A \textbf{%
126}, 259 (1930)

\bibitem{ShelykhJPCM}  I.A. Shelykh, M.A. Kulov, N.G. Galkin and N.T.
Bagraev, \emph{\ J.Phys.:Cond.Matt.} \textbf{19}, 246207 (2007).

\bibitem{key-21}  J. Kondo, \textit{Prog. Theor. Phys.} \textbf{32}, 37
(1964).

\bibitem{Anderson}  P.W. Anderson, \textit{J. Phys. C} \textbf{3}, 2436
(1970).

\bibitem{key-9}  Peter Jaksch, Irina Yamikenko, and Karl-Fredrik Berggren,
\textit{Phys. Rev. B} \textbf{74}, 235320 (2006).

\bibitem{key-36}  A.A. Starikov, I.I. Yakimenko, K.F. Berggren, \emph{Phys.
Rev. B} \textbf{67}, 235319 (2003).

\bibitem{key-37}  K. Hirose, Y. Meir, N. S. Wingreen, Phys. Rev. Lett. 90,
026804 (2003).

\bibitem{key-38}  S. Ihnatenska, I.V. Zozulenko, Phys. Rev. B \textbf{76},
045338 (2007).

\bibitem{key-20}  T. Rejec and Y. Meir, \textit{Nature}, \textbf{442}, 900
(2006)

\bibitem{key-29}  W.D. Oliver, \textit{Ph. D. thesis}, Stanford University,
2003; N. Y. Kim, W. D. Oliver, and Y. Yamamoto, arXiv:0311435

\bibitem{key-30}  P. Roche, J. Segala, and D. C. Glattli, \textit{Phys. Rev.
Lett.} \textbf{93}, 116602 (2004).

\bibitem{key-32}  A. Ramsak and J. H. Jefferson, \textit{Phys. Rev. B}, 71,
161311 (2005)

\bibitem{SplitOFF}  Við skeytum ekki um þriðja orkuborðann sem er langt frá hinum tveimur vegna skiptaverkunar fyrir $k=0$.

\bibitem{note}  Ath. að annað ástand þungu holanna 2hh er venjulega mjög nálægt fyrsta ástandi léttu holanna 1lh og þarf því að taka það með í reikninginn. Hins vegar veldur álag oft til lækkunar á 1lh og því hægt að búa til kerfi þar sem 1lh er nær 1hh en 2hh. Sjá E.L. Ivchenko, G.E. Pikus, Superlattices
and other heterostructures, ISBN 3-540-62030-3, Springer (2003).

\bibitem{SpinFlipHoles}  Spunaskipti af tegundinni$+3/2,-3/2\rightarrow-3/2,+3/2$ hafa alltaf léttar holur sem milliástönd, $+3/2,-3/2\rightarrow+1/2,-1/2\rightarrow-1/2,+1/2\rightarrow-3/2,+3/2$ og falla því líkurnar niður í 0 fyrir stór orkubil $\Delta$.

\bibitem{Estima}  Í stórsæu GaAs er $\kappa \sim -1.2$. Skammtabrunnar með vídd af stærðargráðunni $L\sim 30nm$ hafa Luttinger stuðul $\kappa \sim -0.3$, sjá S.
Glasberg et al., Phys. Rev. B, \textbf{60}, R16295 (1999). Þar af leiðandi gerum við ráð fyrir að $\kappa $ í skammtapunktsnertunni okkar sé þar á milli.
\bibitem{Holes}  M. Rosenau da Costa, I.A. Shelykh, and N.T. Bagraev,
arXiv:0707.1684 (2007).

\bibitem{english}  I. A. Shelykh \textit{et al}, \textit{J. Phys.: Condens. Matter} \textbf{20} 164214 (2008)

\bibitem{DattaAtT} S. Datta (2005), \textit{Quantum Transport: Atom to Transistor}, London: Cambridge University Press

\end{thebibliography}
\end{document}